# Optically induced transport properties of freely suspended semiconductor submicron channels

C. Rossler, K.-D. Hof, S. Manus, S. Ludwig, and J. P. Kotthaus

Department für Physik and Center for NanoScience, Ludwig-Maximilians-Universität, Geschwister-Scholl-Platz 1, D-80539 München, Germany.

J. Simon and A. W. Holleitner

Walter Schottky Institut and Physik Department, Technische Universität München, Am Coulombwall 3, D-85748 Garching, Germany.

D. Schuh and W. Wegscheider

Institut für Experimentelle und Angewandte Physik, Universität Regensburg, D-93040 Regensburg, Germany.

We report on optically induced transport phenomena in freely suspended channels containing a two-dimensional electron gas (2DEG). The submicron devices are fabricated in AlGaAs/GaAs heterostructures by etching techniques. The photoresponse of the devices can be understood in terms of the combination of photogating and a photodoping effect. The hereby enhanced electronic conductance exhibits a time constant in the range of one to ten milliseconds.

Submicron channels and nanowires have for the past few years attracted considerable attention because of the compelling electronic, mechanical and optical properties of low-dimensional systems.[1] Several processes can dominate the photoresponse of such nanosystems[2]-[5], enabling polarization sensitive photonic and optoelectronic devices. For instance, it has been shown that Schottky contacts between nanowires and metal electrodes can give rise to a photoresponse current.[3] In the process, photo-generated electrons and holes are separated and then accelerated by internal electric fields across the semiconductor-metal interface leading to a photoresponse. Furthermore, surface states and adsorbates can govern the photoresponse of semiconductor nanowires by photodesorption at ambient conditions.[4] There, oxygen molecules are adsorbed on the nanowire surface. As negatively charged ions they create a depletion layer with low conductivity near the nanowire surface. Laser excitation gives then rise to photodesorption of the oxygen molecules and consequently to an elevated optically induced conductance. Here, we report on photoconductance experiments which can be understood by a similar photogating effect in combination with a photodoping effect. However, these effects occur at high vacuum conditions. The experiments are performed on freely suspended submicron channels containing a two-dimensional electron gas (2DEG).[6],[7] The devices are fabricated by etching techniques out of an AlGaAs/GaAs heterostructure. Our experimental findings suggest that optically excited electron-hole pairs are spatially separated at the surface of a freely suspended channel due to internal electric fields. On the one hand, trapped excess holes act as a positive local gating voltage (photogating effect). On the other hand, free excess electrons raise the Fermi-energy of the 2DEG throughout the device (photodoping effect). Both effects raise the conductance across the device. The measured photoresponse exhibits a time constant of 1-10 ms, which is consistent with the recombination time of optically induced and spatially separated charge carriers in semiconductor heterostructures.[8] Excess charge carriers are created by interband laser excitation. In that sense, our measurements are complementary to intraband photon-induced current experiments on quantum wires.[9],[10] The latter demonstrated that a far-infrared excitation of quantum wires can give rise to heating and rectification processes of electrons in the 2DEG.[11],[12] We can exclude such processes by measuring the photoresponse as a function of the source-drain voltage and by spatially mapping the photoresponse of the freely suspended channels and the adjacent electrodes.

Starting point is a modulation-doped GaAs/AlGaAs heterostructure with a thickness of 130 nm (active layer) on top of 400 nm AlAs (sacrificial layer).[6],[7] 40 nm below the surface of the heterostructure, a 25 nm thick GaAs-quantum well contains a 2DEG with an electron sheet density of $n_S \sim 5.5 \cdot 10^{11} \, \text{cm}^{-2}$ and electron mobility of $\mu \sim 7.8 \cdot 10^5 \, \text{cm}^2/\text{Vs}$. The freely suspended channels are created by the following steps. First, a protective nickel layer with a thickness of 60 nm is defined at specific areas of the surface by means of optical lithography (outer leads) and e-beam lithography (center region). Then, anisotropic reactive ion etching with SiCl$_4$ is employed to completely remove the unprotected top layers of the heterostructure. The nickel layer is removed by FeCl$_3$. Finally, the sacrificial layer is removed by isotropic wet-etching utilizing 1% hydrofluoric acid. Hereby, the active layer is undercut edgeways and freely suspended channels can be defined, which contain a 2DEG.[7] The SEM-graph (Fig. 1a) shows the sample surface, where the undercut regions (pale areas with a width of approximately 2 μm) frame the active layer. In the following, the particular channel of Fig. 1b is characterized. It has dimensions of about 4 μm x 600 nm x 130 nm (length x width x height). The width at the narrowest part of the channel is chosen to be wide enough (350 nm), so that the freely suspended channel can be still regarded as quasi two-dimensional.[6],[13],[14] The conductance across the channel is measured from source to drain (Fig. 1a), and all electrical leads are contacted by annealed pads made of AuGeNi. The electronic width of the channel can be controlled by applying voltages to the side gates G1 and G2, while region G3 is electronically connected to the drain contact by a small bridge. All measurements are carried out in a helium continuous-flow cryostat at a vacuum of about $10^{-5}$ mbar and a bath temperature of 3.5 K. Charge carriers are locally excited by focusing the light of a mode-locked titanium:sapphire laser with a repetition rate of 76 MHz through the objective of a microscope onto the surface of the sample. With a spot diameter of 2 μm the power density is chosen to be 3 mW/cm$^2$ at a photon energy $E_{PHOTON}$ = 1.55 eV, if not stated otherwise. We find that the channel can be pinched off at $V_{G1}$ ~ -42 V at low temperatures ($T_{2DEG}$ ~ 10 K). Fig 1c shows the source-drain current $I_{SD}$ as a function of the source-drain voltage $V_{SD}$ at $V_{G1}$ = -20 V. The dark current $|I_{OFF}|$ (solid line) clearly increases, when the freely suspended channel is illuminated (dashed line). For the photoresponse measurements, we chop the laser at frequency $f_{CHOP}$. The resulting ac-photoresponse $|I_{PR}(E_{PHOTON}, f_{CHOP})| = |I_{ON} - I_{OFF}|$ $E_{PHOTON}, f_{CHOP}$ across the sample with the laser being "ON" or "OFF", respectively, is amplified by a current-voltage converter and detected with a lock-in amplifier utilizing the reference signal provided by the chopper. Fig. 1d depicts such a photoresponse measurement as a function of $V_{SD}$, while the channel is illuminated. We observe that the offset of $V_{SD}^{off} \sim -40\,\mu V$ at $|I_{PR}| = 0$ does not depend on the laser power. Hereby, we can deduce that rectification or thermopower-effects do not dominate the conductive photoresponse in our device.[10],[11],[12] Instead, we find that the input-offset is given by the current-voltage amplifier. In Fig. 1e, the spatially resolved photoresponse of the area shown in Fig. 1a is depicted. $|I_{PR}|$ is measured, while the position of the sample is moved in the x-y-plane in steps of 1 µm. The measured photoresponse ranges from $|I_{PR}| \sim 30$ pA, when illuminating the GaAs substrate, $|I_{PR}| \sim 300$ pA, when shining light onto the active layer ($V_{SD} = 1$ mV, $f_{CHOP} = 73$ Hz) to a maximum of $|I_{PR}| \sim 2.8$ nA at the position of the freely suspended channel. When shining light onto the biased gate G1 (see dashed lines in Fig. 1e), $|I_{PR}|$ is suppressed by one order of magnitude as compared to illuminating the areas G2 or G3. This variation excludes global heating effects due to the laser irradiation as the major cause of the photoresponse.[11] We interpret our observations by a combination of a photogating effect and a photodoping effect, as discussed below.

Fig. 2a depicts the photoresponse of a second freely suspended channel as a function of the photon energy. The second sample is fabricated out of the same wafer as the first sample with device details described in [15]. The sigmoid fit to the photoresponse is centered at a photon energy of $E_{PR} \sim 1.535$ eV which is larger than the band gap energy of GaAs ($E_{GaAs} \sim 1.52$ eV) at T = 4 K and much smaller than $E_{AlGaAs} \sim 1.90$ eV. This indicates that the photoresponse is caused by excitons created in the GaAs quantum well and not in the GaAs substrate or the AlGaAs layers. In Fig. 2b, the photoluminescence spectrum of the quantum well at an undercut (non-undercut) region is represented by filled (open) squares. The maxima have an energy of $E_{UC} = 1.545$ eV and $E_{NUC}$ = 1.550 eV. We interpret the observed red shift of $E_{PR}$ and $E_{UC}$ as compared to $E_{NUC}$ by the presence of an electric field at the undercut regions and the resulting Stark-shift.[16] The origin of the electric field can be attributed to the mid-gap Fermi-level pinning of the surface states of GaAs with respect to the electrostatic potential of the 2DEG.[17] The expected field strength can be estimated by the band gap energy of GaAs and the lateral depletion length of typically $l_{DEP}$ ~ 100 nm[18]. The resulting field strength $F$ ~ $0.5 \cdot E_{GaAs}$ / (e $\cdot$ $l_{DEP}$) = 7.6·10$^6$ V/m is sufficient to ionize optically created excitons in the undercut regions of the quantum well, and to explain the red-shift of $E_{PR}$ and $E_{UC}$ as compared to $E_{NUC}$ (Fig. 2a,b).[16],[19]

Separated charge carriers can contribute twofold to the photoresponse of the device. On the one hand, the holes can drift to the edge of the nanostructure, acting as a local positive gating voltage (photogating effect). On the other hand, the electrons can increase the electron sheet density of the 2DEG; raising the Fermi-energy within all electronically connected areas of the active layer (photodoping effect).[20],[21] Both effects result in an optically increased conductance of the device and they can be distinguished by their spatial occurrence in Fig. 1e. The local photogating effect is most pronounced at the narrowest conducting junction of the device (i.e. the bridge) whereas the photodoping effect acts globally without the need to illuminate the freely suspended channel itself. Generally, both the photogating and the photodoping effect also explain the linear dependence of the photoresponse as a function of $V_{SD}$ in Fig. 1d, which corresponds to a change in conductance of $\Delta G \approx 2 \mu S$.

Spatially separated holes and electrons typically recombine within milliseconds to seconds.[8] In order to test the time scale of the recombination, we measure the photoresponse of the first sample as a function of the chopper frequency (Fig. 3a). The double-logarithmic representation of the data reveals a corresponding time constant in the range of 1-10 ms for both the freely suspended channel (squares) and the source-drain leads (open squares). Such a slow time constant substantiates the interpretation that the photoresponse is dominated by spatially separated charge carriers. The design of the first sample allows testing our interpretation further: region G3 is directly connected to drain (Fig. 1a), while gate G2 is connected to drain on-chip at the annealed AuGeNi pads. Both regions show a photoresponse with a slow time constant (Fig. 1e and 3a). In contrast, the biased gate G1 is not electrically connected to the source-drain region and therefore, illumination of G1 should not result in a global photodoping effect. Consistently, the illumination of gate G1 does not result in a photoresponse above the experimental noise level (see dashed lines in Fig. 1e).

We would like to note that a bolometric response would have a similarly slow time constant.[23] However, a bolometric effect alone cannot explain the homogeneous photoresponse at regions G2 and G3 in Fig. 1e. Finally, Fig. 3b demonstrates that all measurements are taken in a linear response regime; i.e. the photoresponse increases linearly for laser intensities $P_{LASER}$ < 16 mW/cm$^{-2}$. Again, this finding is consistent with the interpretation in terms of dominating photogating and photodoping effects.

In summary, we present spatially and spectrally resolved photoresponse measurements and photoluminescence measurements on freely suspended submicron channels. We interpret the results in terms of the combination of a photogating and a photodoping effect, i. e. the consequence of the spatial separation of holes and electrons.

The measured photoresponse exhibits a time constant of 1-10 ms, which is consistent with the recombination time of optically induced and spatially separated charge carriers.

We gratefully acknowledge financial support from BMBF via nanoQUIT, the DFG (Ho 3324/4) and the German excellence initiative via the "Nanosystems Initiative Munich (NIM)".

Fig. 1(a) (color online): Scanning electron micrograph of the device. The active layer resides on a socket which is edgewise undercut (pale stripes). Submicron sized channels are fully suspended. (b) Side view of the center region under a tilt angle of 75 degree. (c) Source-drain current $I_{SD}$ as a function of source-drain voltage $V_{SD}$ (solid line) and with illumination of the channel (dashed line) ($T_{2DEG}$ ~10 K and $V_{G1}$ = -20.5 V). (d) Corresponding absolute value of the photoresponse $|I_{PR}|$ ($f_{CHOP}$ = 74 Hz). (e) Spatially resolved photoresponse $|I_{PR}|$ ($V_{G1}$ = -20.5 V, $V_{SD}$ = 1 mV, $T_{2DEG}$ ~ 10 K, $P_{LASER}$ ~ 3 mW/cm2 and $f_{CHOP}$ = 72.5 Hz). The scanned area is identical to the one in (a), with the contour of the biased Gate G1 shown as dashed white lines as a guide to the eyes. See text for details.

Fig. 2(a) Photoresponse $|I_{PR}|$ of the second sample as a function of excitation photon energy $E_{PHOTON}$. The sigmoid fit is a guide to the eye centered at $E_{PR}$ = 1.535 eV. (b) Spectrally resolved photoluminescence $PL$ of the quantum well at an undercut region (filled squares with a lorentzian fit centered at $E_{UC}$ = 1.545 eV) and at a non-undercut region (open squares with a lorentzian fit centered at $E_{NUC}$ = 1.550 eV). Data taken in a microPL setup at $T_{2DEG}$ ~ 10 K.

Fig. 3 (a) Photoresponse $|I_{PR}|$ of the channel (filled squares) and of the source-lead (open squares) of the first sample as a function of the chopping frequency $f_{CHOP}$ in a double-logarithmic representation at 6 mW/cm$^2$. (b) Photoresponse $|I_{PR}|$ of the second sample as a function of laser excitation power taken at $E_{PHOTON}$ = 1. 653 eV and $f_{CHOP}$ = 116 Hz. All data are taken at $T_{2DEG}$~ 10 K.

References:

[1] See for example, M. Roukes, "Nanoelectromechanical Systems" in "Technical Digest of the 2000 Solid State Sensor and Actuator Workshop" (Transducers Research Foundation, Cleveland, OH; 2000) ISBN 0-9640024-3-4, arXiv:cond-mat/0008187 (2000); and J. Hu, T.W. Odom, C.M. Lieber, Acc. Chem, Res. **32**, 456 (1999).
[2] J. Wang, M.S. Gudiksen, X. Duan, Y. Cui, C.M. Lieber, Science **293**, 1455 (2001).
[3] Y. Gu, E.-S. Kwak, J.L. Lensch, J.E. Allen. T.W. Odom, L. J. Lauhon, Appl. Phys. Lett. **87**, 043111 (2005).
[4] H. Kind, H. Yan, B. Messer, M. Law P. Yang, Adv. Mat. **14**, 158 (2002).
[5] H. Pettersson, J. Tragardh, A. I. Persson, L. Landin, D. Hessman, and L. Samuelson, Nano Lett. **2**, 229 (2006).
[6] J. Kirschbaum, E. M. Hohberger, R. H. Blick, W. Wegscheider, and M. Bichler, Appl. Phys. Lett. **81**, 280 (2002).
[7] C. Rossler et al. Nanotechnology **19**,165201 (2008).
[8] S. Zimmermann, A. Wixforth, J.P. Kotthaus, W. Wegscheider, M. Bichler, Science **283**, 1292 (1999).
[9] Q. Hu, Appl. Phys. Lett. **62**, 837 (1993).
[10] Q. Hu, S. Verghese, R. A. Wyss, T. Schäpers, J. del Alamo, S. Feng, K. Yakubo, M. J. Rooks, M. R. Melloch, and A. Förster, Semicond. Sci. Technol. **11**, 1888-1894 (1996).
[11] R. Wyss, C. C. Eugster, J. A. del Alamo, and Q. Hu, Appl. Phys. Lett. **63**, 1522 (1993).
[12] T. J. B. M. Janssen, J. C. Maan, J. Singleton, N. K. Patel, M. Pepper, J. E. F. Frost, D. A. Ritchie, and G. A. C. Jones, J. Phys.: Condens. Matter 6, L163 (1994).
[13] Generally, no conductance steps of $2e^2/h$ are detected for the presented channels. Such steps would demonstrate that a one-dimensional quantum point contact is defined within the freely suspended channel. E.g. D.A.Wharam, et al. J. Phys. C **21**, L209 (1988); B.J. van Wees, et al. Phys. Rev. Lett. **60**, 848 (1988).
[14] A. W. Holleitner, V. Sih, R. C. Myers, A. C. Gossard, D. D. Awschalom, Phys. Rev. Lett. **97**, 036805 (2006).
[15] K.-D. Hof, C. Rossler, W. Wegscheider, S. Ludwig, and A. W. Holleitner, Physica E **40**, 1739-1741 (2007).
[16] D.A.B. Miller et al., Phys. Rev. B **32**, 2, 1043 (1985).
[17] A. Gärtner, L. Prechtel, D. Schuh, A. W. Holleitner, and J. P. Kotthaus, Phys. Rev. B **76**, 085304 (2007).
[18] T.J. Thornton et al. Phys. Rev. Lett. **63**, 2128 (1989).
[19] A. Schmeller et al. Appl. Phys. Lett. **64**, 330 (1993).
[20] T. Ando, A.B. Fowler, F. Stern, Rev. Mod. Phys. **54**, 437 (1982).
[21] The magnitude of the observed photoresponse of $|I_{PR}| \sim 300\,\text{pA}$ (Fig. 1e) suggests a relative change in the number of electrons per laser chopping cycle

$\Delta N_e / N_e \propto \Delta E_F / E_F \propto |I_{PR}|/|I_{SD}| \cong 300 \text{ pA}/200 \text{ nA} \cong 10^{-3}$, with $E_F$ the Fermi-energy, while the total number of electrons in the active layer can be estimated to be $N_e = A_{active} \cdot n_s \cong 2 \cdot 10^{-3} \text{ cm}^2 \cdot 5 \cdot 10^{11} \text{ cm}^{-2} = 10^9$, with $n_s$ the electron sheet density and $A_{active}$ the active area of the 2DEG. The corresponding number of excess electrons $\Delta N_e = 10^6$ is smaller than the number $N_{PHOTON}$ of incident photons per cycle $N_{PHOTON} = P_{LASER}/(E_{PHOTON} \cdot 2 \cdot f_{CHOP})$ $= 300 \text{ pW}/(1.55 \text{ eV} \cdot 2 \cdot 74 \text{ Hz}) \approx 10^7$, with $P_{LASER}$ the laser power; thus enabling the discussed photodoping effect.

[22] L. Schultheis et al. Phys. Rev. B **34**, 12, 9027 (1986).
[23] F. Neppl, J.P. Kotthaus, J. F. Koch, Phys. Rev. B **19**, 5240 (1979).

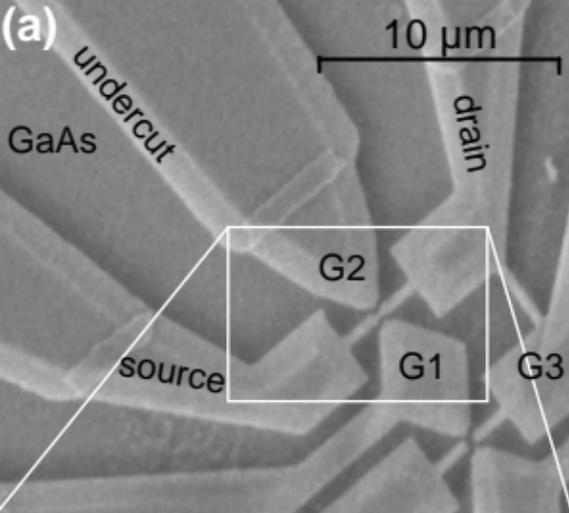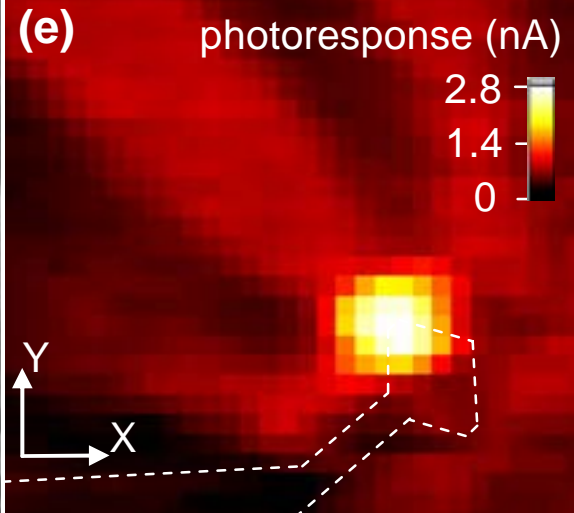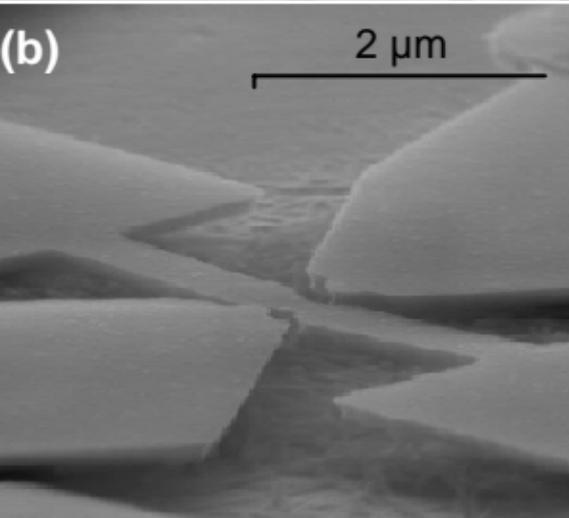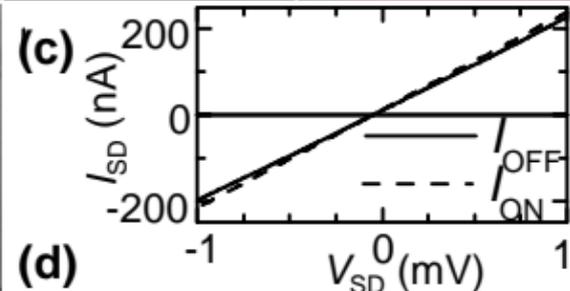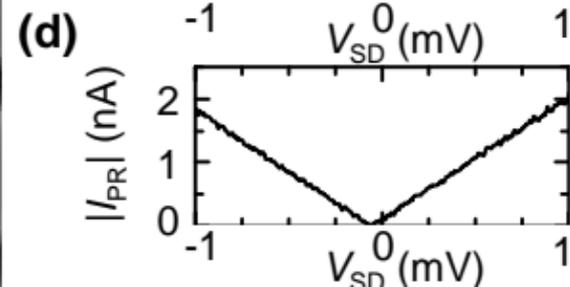

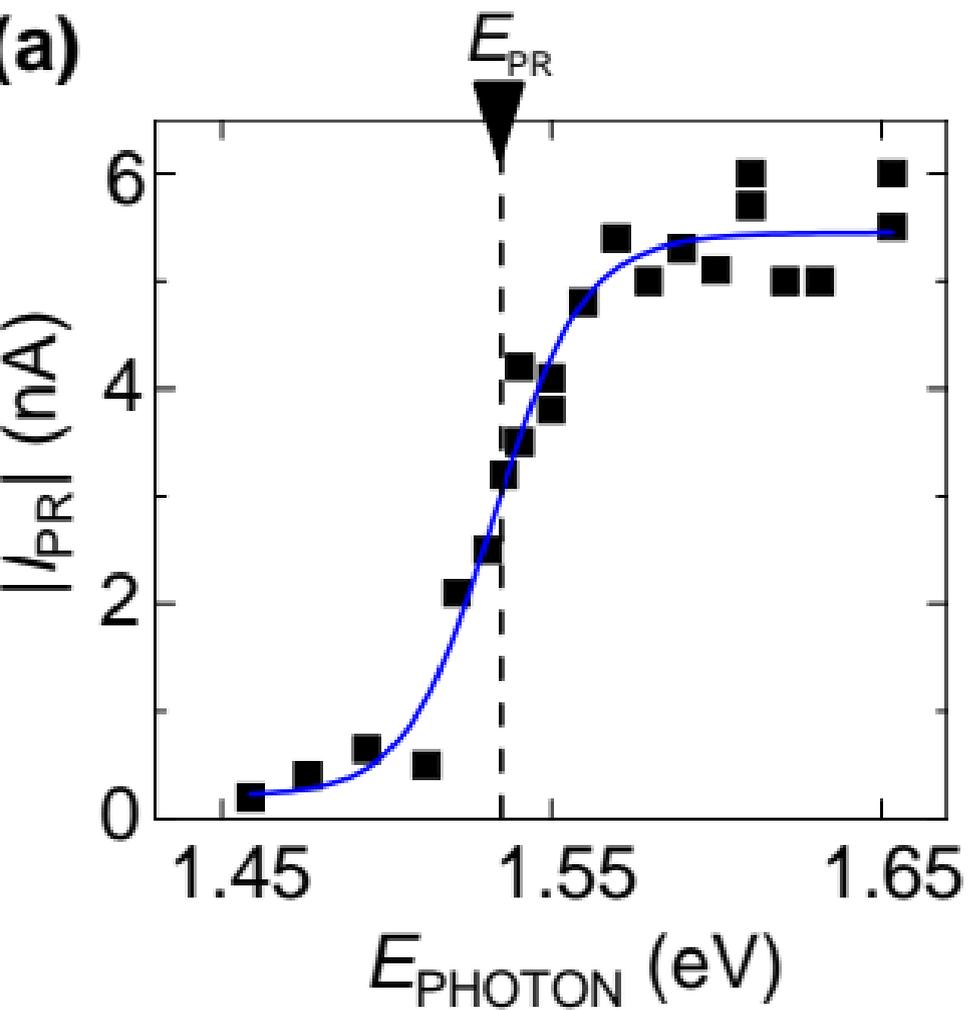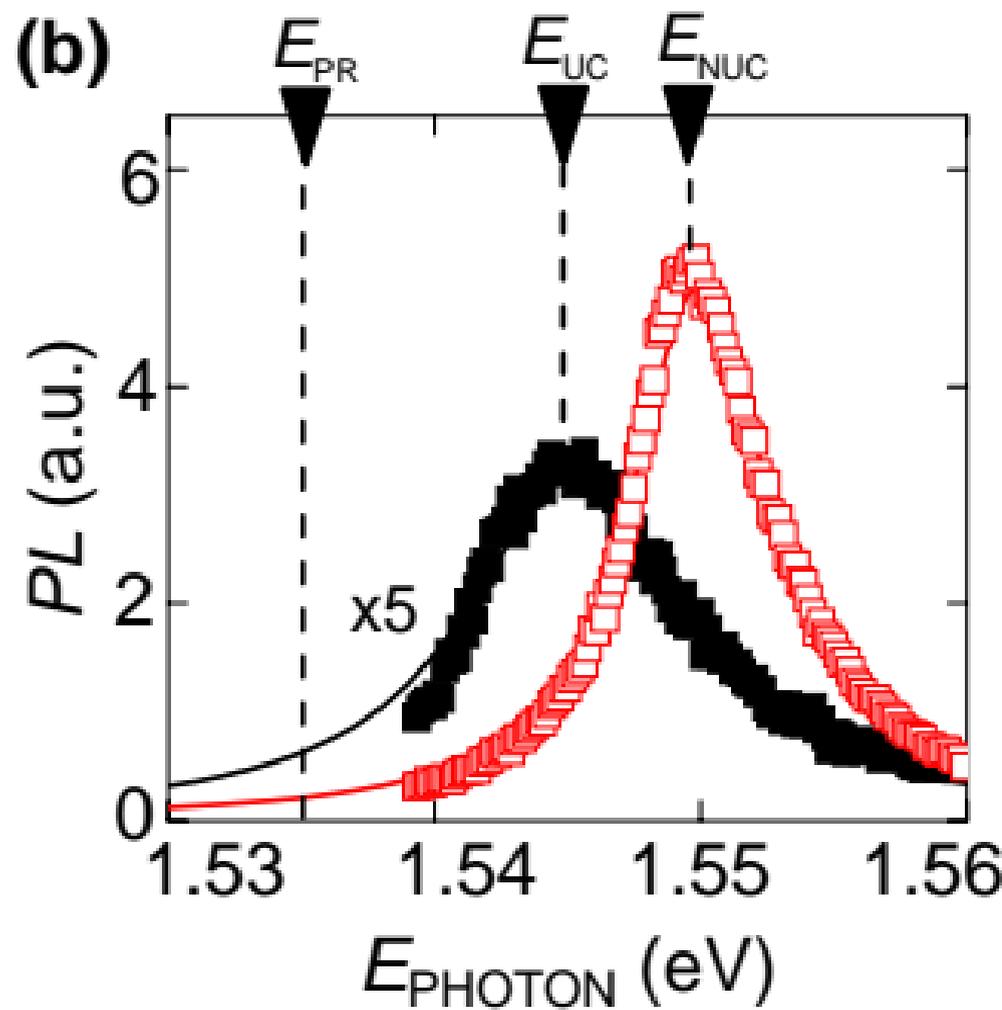

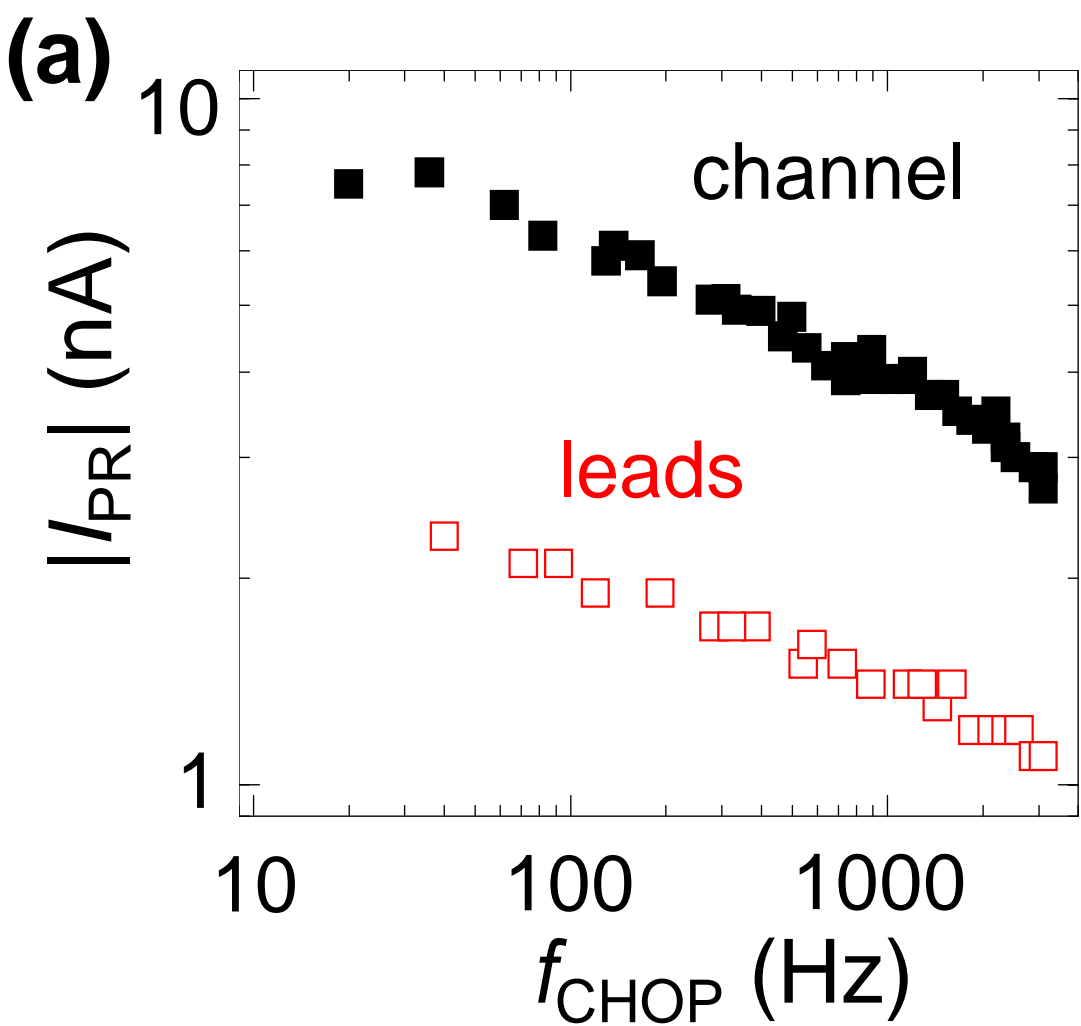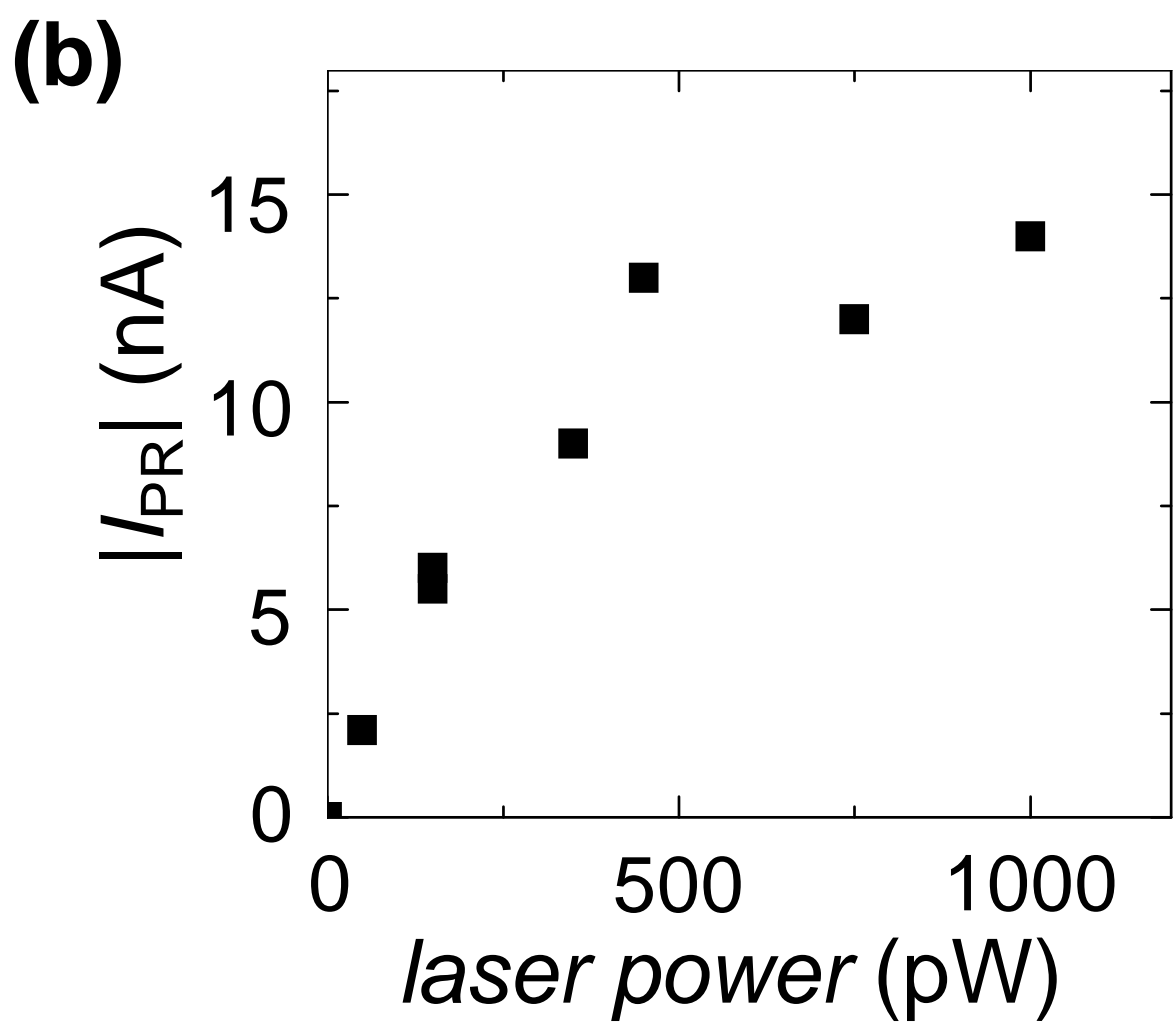